\documentclass[11pt,american,twoside,a4wide]{article}
\usepackage[T1]{fontenc}
\usepackage[latin1]{inputenc}
\usepackage{latexsym}
\usepackage{amsmath}
\usepackage{bbm}
\usepackage{amssymb}
\usepackage{babel}
\usepackage{amsfonts}
\usepackage{times}
\usepackage{theorem}
\usepackage{epsfig}
\usepackage{enumerate}
\usepackage{color}
\usepackage[active]{srcltx}
\usepackage[colorlinks=true]{hyperref}
\usepackage{fancyhdr}

\numberwithin{equation}{section}

\newcounter{smallarabics}

\newcounter{smallroman}

\newcommand{\ben}{\begin{enumerate}[{\rm (1)}]}
\newcommand{\een}{\end{enumerate}}

\newtheorem{theoreme}{Theorem}[section]
\newtheorem{proposition}[theoreme]{Proposition}
\newtheorem{lemma}[theoreme]{Lemma}
\newtheorem{definition}[theoreme]{Definition}

\def\rr{{\mathbb R}}
\def\zz{{\mathbb Z}}
\def\cc{{\mathbb C}}

\def\textsl{{}}

\def\Im{{\rm Im}\,}

\def\cT{\mathcal{T}}

\def\c0inf{C_0^\infty}
\def\bep{\begin{proposition}}
\def\eep{\end{proposition}}

\def\proof{\noindent {\bf Proof.}\ \ }

\def\cH{{\cal  H}}

\newcommand{\bra}{\langle} 
\newcommand{\ket}{\rangle}

\def\per{{\rm per}}

\def\i{{\rm i}}
\newcommand{\beq}{\begin{equation}}
\newcommand{\eeq}{\end{equation}}
\newcommand{\bear}[1]{\begin{array}{#1}}
\newcommand{\ear}{\end{array}}

\def\sp{{\hat e}}

\newcommand{\e}{\mathrm{e}}
\renewcommand{\i}{\mathrm{i}}

\renewcommand{\d}{\mathrm{d}}

\def\qed{$\Box$\medskip}

\def\cT{{\cal T}}

\def\bel{\begin{lemma}}
\def\eel{\end{lemma}}
\def\bet{\begin{theoreme}}
\def\eet{\end{theoreme}}
\def\bed{\begin{definition}}
\def\eed{\end{definition}}

\def\12{\frac{1}{2}}

\def\e{{\rm e}}

\def\d{{\rm d}}

\def\cH{{\cal H}}

\def\ac{{\rm ac}}

\def\sp{{\rm sp}}
\def\cS{{\cal S}}

\def\tr{{\rm tr}}

\def\bra{\langle}
\def\ket{\rangle}

\def\per{{\rm per}}

\newcommand{\ds}{\displaystyle}
\begin{document}
\def\today{}
\title{Crystalline conductance and absolutely continuous spectrum\\ of  1D samples}
\author{L. Bruneau$^{1}$, V. Jak\v{s}i\'c$^{2}$, Y. Last$^3$, C.-A. Pillet$^4$
\\ \\ 
$^1$ D\'epartement de Math\'ematiques and UMR 8088\\
CNRS and Universit\'e de Cergy-Pontoise\\
95000 Cergy-Pontoise, France
\\ \\
$^2$Department of Mathematics and Statistics\\ 
McGill University\\
805 Sherbrooke Street West \\
Montreal,  QC,  H3A 2K6, Canada
\\ \\
$^3$Institute of Mathematics\\
The Hebrew University\\
91904 Jerusalem, Israel
\\ \\
$^4$Universit\'e de Toulon, CNRS, CPT, UMR 7332, 83957 La Garde, France\\
Aix-Marseille Universit\'e, CNRS, CPT, UMR 7332, Case 907, 13288 Marseille, France\\
FRUMAM
}
\maketitle
\thispagestyle{empty}
\begin{quote}
\noindent{\bf Abstract.} We characterize the absolutely continuous spectrum
of half-line one-dimensional Schr\"odinger operators in terms of the
limiting behavior of the Crystaline Landauer-B\"uttiker conductance of the 
associated finite samples. 
\end{quote}

\section{Introduction} 

This note is a direct continuation of~\cite{BJLP2} and completes the research 
program initiated in~\cite{BJP}. This program concerns characterization of 
the absolutely continuous spectrum of half-line discrete Schr\"odinger 
operators in terms of the limiting conductances of the associated finite 
samples and is intimately linked with the celebrated Schr\"odinger 
Conjecture/Property of Schr\"odinger operators~\cite{Av,MMG}. 
In~\cite{BJLP2}, this characterization was carried out for the well-known  
Landauer-B\"uttiker (LB) and the Thouless (Th) conductance. Here we extend 
these results to the Crystaline Landauer-B\"uttiker (CLB) conductance 
introduced in~\cite{BJLP1}. The CLB conductance provides a natural link 
between the LB and Th conductances and is likely to play an important role 
in future studies of transport properties of 1D samples. 

We briefly recall the setup and results of~\cite{BJLP1, BJLP2}, referring the 
reader to the introductions of these papers for details and additional 
information. The starting point is the discrete Schr\"odinger 
operator\footnote{The setup and all our results extend to the case of 
Jacobi matrices, see~\cite{BJLP3}. For notational simplicity we will restrict 
ourselves here to the physically relevant case of Schr\"odinger operators.}
\[
h_\cS=-\Delta + v,
\]
acting on the Hilbert space $\ell^2(\zz_+)$, where $\zz_+$ denotes the set of 
positive integers.\footnote{For our purposes, the choice of boundary condition 
is irrelevant. For definiteness we shall impose Dirichlet b.c. on the discrete 
Laplacian $\Delta$.} The operator 
$h_\cS$ is the one-electron  Hamiltonian of the extended sample. The 
one-electron Hamiltonian $h_{\cS_L}$ of the sample of length $L$ is obtained 
by restricting $h_\cS$ to ${\cal H}_L=\ell^2(\zz_L)$, where 
$\zz_L=\{1, \cdots, L\}$. In the Electronic Black Box (EBB) model considered 
in~\cite{BJLP1,BJLP2}, this finite sample is connected at its end points $1$ 
and $L$ to reservoirs described by the following one-electron data: Hilbert 
spaces $\cH_{l/r}$, where $l/r$ stands for left/right, Hamiltonians $h_{l/r}$, 
and unit vectors $\psi_{l/r}\in\cH_{l/r}$. For latter reference, we introduce 
the functions
\beq\label{reserv-funct}
F_{l/r}(E)=\langle\psi_{l/r},(h_{l/r}-E-\i0)^{-1}\psi_{l/r}\rangle,
\eeq
and the sets (note that $\Im F_{l/r}(E)\ge0$) 
\[
\Sigma_{l/r}=\{E\,:\,\Im F_{l/r}(E)>0\}.
\]
In the absence of coupling, the one-electron Hamiltonian of the joint  system 
sample+reservoirs is 
$$
h_{0,L}=h_l + h_{\cS_L}+h_r,
$$
acting on $\cH=\cH_l\oplus \cH_L\oplus \cH_r$. The junctions between the 
sample and the reservoirs are described by 
$$
h_{T,l}=|\psi_l\ket\bra\delta_1|+|\delta_1\ket\bra \psi_l|\qquad
\mbox{and}\qquad
h_{T,r}=|\psi_r\ket\bra\delta_L|+|\delta_L\ket\bra\psi_r|,
$$
and the coupled one-electron Hamiltonian is
$$
h_{\kappa,L}=h_{0,L}+\kappa(h_{T,l}+h_{T,r}), 
$$
where $\kappa\not=0$ is a coupling constant. The left/right reservoir is  
initially at equilibrium at zero temperature and chemical potential 
$\mu_{l/r}$ where $\mu_l<\mu_r$. The voltage differential induces a steady 
state charge current from the right to the left reservoir across 
the sample and the corresponding conductance  is given by the 
Landauer-B\"uttiker formula, see, e.g., \cite{La,BILP,AJPP,CJM,N},
\begin{equation}\label{eq:lbformula}
G_{{\rm LB}}(L,I)=\frac1{2\pi|I|}\int_I\cT_{\rm LB}(L,E)\,\d E,
\end{equation}
where $I=(\mu_l,\mu_r)$, $|I|=\mu_r-\mu_l$, and
\begin{equation}\label{eq:transmissionproba}
\cT_{\rm LB}(L,E)= 4\kappa^4|\langle\delta_1,(h_{\kappa,L}-E-\i0)^{-1}\delta_L\rangle|^2\,
\Im F_l(E)\,\Im F_r(E),
\end{equation}
is the transmission probability from the right to the left reservoir at energy 
$E$. Obviously, only the energies in $\Sigma_l \cap \Sigma_r$ contribute to 
the integral~\eqref{eq:lbformula}. For this reason, in some applications of 
the LB formula we will need to assume the {\em transparency condition} 
that $I\subset\Sigma_l\cap\Sigma_r$ (see Theorems~\ref{main-th} 
and~\ref{main-th-1}).

The Thouless formula is the special case of Landauer-B\"uttiker formula 
in which the reservoirs are implemented in such a way that the coupled 
Hamiltonian is the periodic discrete Schr\"odinger operator on $\ell^2(\zz)$ 
\[
h_{\kappa, L}=h_{{\rm per},L}=-\Delta+v_{{\rm per},L},
\]
where the sample potential $v(n)$ is extended from $\zz_L$ to $\zz$ by 
setting $v_{\per,L}(n+mL)=v(n)$ for $n\in \zz_L$ and $m\in \zz$. We shall 
refer to the corresponding EBB model as {\em crystaline}  (see~\cite{BJLP1} 
for details). In this case the transport is reflectionless and the
Landauer-B\"uttiker formula coincides  with the  Thouless formula:
\beq
G_{\rm Th}(L,I)=\frac{|\sp(h_{{\rm per},L})\cap I|}{2\pi |I|},
\label{thouless-heuri-2}
\eeq
where $\sp(h_{\per,L})$ denotes the spectrum of $h_{{\rm per},L}$. 

\begin{figure}
\centering
\includegraphics[scale=0.4]{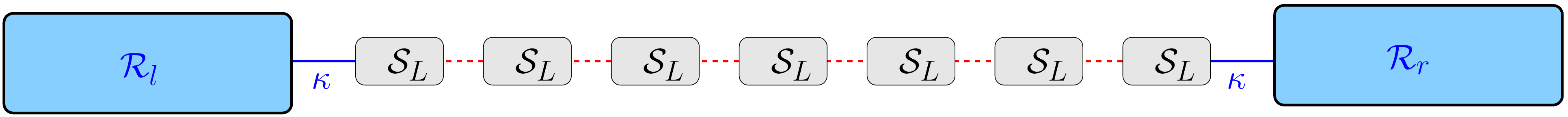}
\caption{The EBBM described by the Hamiltonian $h_{\kappa,L}^{(N)}$ for $N=7$.}
\label{Fig1}
\end{figure}

The CLB conductance has appeared implicitly in early physicists works on 
Thouless conductance~\cite{ET}. The following precise mathematical definition 
was proposed in~\cite{BJLP2}. Consider the Landauer-B\"uttiker formula 
$G_{\rm LB}^{(N)}(L,I)$ for the model in which the sample $\cS_L$ is replaced 
by its $N$-fold repetition while the reservoirs remain fixed (see 
Figure~\ref{Fig1}). 
The limit $N\to\infty$ then gives the CLB formula. To describe it, let
$h_{{\rm per},L}^{(l)}$ and $h_{{\rm per},L}^{(r)}$ be the restrictions of 
$h_{{\rm per},L}$ to $\ell^2((-\infty,0]\cap\zz)$  and 
$\ell^{2}([1,\infty)\cap\zz)$ with Dirichlet boundary conditions, and 
\beq
\begin{split}
m_l(L,E)&=\langle\delta_{0},(h_{{\rm per},L}^{(l)}-E-\i0)^{-1}\delta_{0}\rangle,\\[3mm]
m_r(L,E)&=\langle\delta_{1},(h_{{\rm per},L}^{(r)}-E-\i0)^{-1}\delta_{1}\rangle.
\end{split}
\label{Mfunct}
\eeq
We set
$\cT_{\rm CLB}(E)=0$ for 
$E\in\rr\setminus(\sp(h_{{\rm per},L})\cap\Sigma_l\cap\Sigma_r)$ and 
\beq
\cT_{\rm CLB}(L,E)=\left[1+\frac14\left(
\frac{|m_r(L,E)-\kappa^2F_r(E)|^2}
{\Im(m_r(L,E))\Im(\kappa^2F_r(E))}
+\frac{|m_l(L,E)-\kappa^2F_l(E)|^2}
{\Im(m_l(L,E))\Im(\kappa^2F_l(E))}
\right)\right]^{-1}
\label{def:crys-tra-coef}
\eeq
for $E\in\sp(h_{{\rm per},L})\cap\Sigma_l\cap\Sigma_r$. The Crystaline Landauer-B\"uttiker conductance is defined as
\[
G_{\rm CLB}(L,I)=\frac{1}{2\pi |I|}\int_I{\cal T}_{\rm CLB}(L,E)\d E.
\]
In~\cite{BJLP1} it was shown that
\[
\lim_{N\to\infty}G_{\rm LB}^{(N)}(L,I)= G_{\rm CLB}(L,I),
\]
and that 
\beq\label{eq:optimal}
G_{\rm Th}(L,I)=\sup G_{\rm CLB}(L,I),
\eeq
where the supremum is taken over all realizations of the reservoirs. 
The latter identity 
clarifies the common heuristics in the physics literature that the Thouless 
conductance should be considered as an upper bound on the possible 
conductances of a finite  sample. Since the supremum  is achieved precisely for 
the crystaline EBB model \cite{BJLP1}, it also  identifies the  heuristic notion of 
``optimal feeding'' of the sample by reservoirs, needed to reach the Thouless conductance, with the reflectionless electron  transport across junctions.

Let $\sp_{\rm ac}(h_\cS)$ denote the absolutely continuous spectrum of $h_\cS$. The main result of~\cite{BJLP2} is:

\bet\label{main-th} 
Let $(L_k)$ be  a sequence of  positive integers satisfying $\lim L_k=\infty$. Consider the following statements:
\begin{enumerate}[{\rm (1)}]
\item $$I \cap \sp_\ac(h_\cS)=\emptyset.$$
\item 
\[\lim_{k\rightarrow \infty}G_{\rm LB}(L_k, I)=0.
\]
\item 
\[\lim_{k\rightarrow \infty}G_{\rm Th}(L_k, I)=0.
\]
\end{enumerate}
Then $(1)\Rightarrow(2)$ and $(1)\Leftrightarrow(3)$. If the transparency 
condition $I\subset\Sigma_l\cap\Sigma_r$ holds, then also 
$(2)\Rightarrow(1)$, and all three statements are equivalent.
\eet
{\bf Remark.} The transparency condition is necessary for the implication 
$(2)\Rightarrow(1)$, and in our context it should be considered as an 
assumption on  the non-triviality of the setup. The same remark applies to 
Theorem~\ref{main-th-1} below. 

In this note we complete Theorem~\ref{main-th} with: 

\bet\label{main-th-1} Let  $(L_k)$ be a sequence  of positive integers satisfying $\lim L_k=\infty$. Consider the following statements:
\begin{enumerate}[{\rm (1)}]
\item
$$
I\cap \sp_\ac(h_\cS)=\emptyset.
$$
\item 
\[
\lim_{k\rightarrow \infty}G_{\rm CLB}(L_k, I)=0.
\]
\end{enumerate}
Then $(1)\Rightarrow(2)$. If the transparency condition 
$I\subset\Sigma_l\cap\Sigma_r$ holds, then also $(2)\Rightarrow(1)$.
\eet

Theorems~\ref{main-th} and~\ref{main-th-1} naturally lead to questions 
regarding the relative scaling and the rate of convergence to zero  
of the conductances $G_{\#}(L,I)$, $\#\in\{{\rm LB},{\rm Th},{\rm CLB}\}$,
in the regime $I\cap\sp_{\rm ac}(h_\cS)=\emptyset$. Although these questions
played a prominent role in early physicists works on the subject (see, e.g.,
\cite{AL,CGM}) we are not aware of any mathematically rigorous works on this 
topic. We plan to address these problems in the continuation of our research 
program. 

\bigskip\noindent
{\bf Acknowledgment.} The research of V.J. was partly supported by NSERC. The 
research of Y.L.\ was partly supported by The Israel Science Foundation
(Grant No.\;1105/10) and by Grant No.\;2014337 from the United States-Israel
Binational Science Foundation (BSF), Jerusalem, Israel. A part 
of this work has been done during a visit of L.B. to McGill University 
supported by NSERC. The work of C.-A.P. has been carried out in the framework 
of the Labex Archim\`ede (ANR-11-LABX-0033) and of the A*MIDEX project 
(ANR-11-IDEX-0001-02), funded by the ``Investissements d'Avenir'' French 
Government program managed by the French National Research Agency (ANR).

\section{Proofs}\label{sec:proof}

\subsection{Proof of Theorem \ref{main-th-1}} 

Our proof proceeds by showing that for any sequence $(L_k)$, the vanishing of 
the CLB conductance is equivalent to the vanishing of the Th conductance.
Due to~\eqref{eq:optimal}, it suffices to prove that if 
$I\subset\Sigma_l\cap\Sigma_r$, then 
\beq\label{eq:crys-to-thouless}
\lim_{k\rightarrow \infty}G_{\rm CLB}(L_k,I)=0
\quad \Longrightarrow \quad 
\lim_{k\rightarrow \infty}G_{\rm Th}(L_k,I)=0.
\eeq

Relation~\eqref{def:crys-tra-coef} expresses the CLB conductance in terms of 
the $m$-functions $m_l$ and $m_r$ of the periodic operator $h_{{\rm per},L}$. 
The first part of the proof consists in estimating these $m$-functions in 
terms of the norm of the transfer matrix 
$$
T(L,E)=\left[\begin{matrix}
v(L)-E&-1\\
1&0
\end{matrix}\right]
\cdots\left[\begin{matrix}
v(1)-E&-1\\
1&0
\end{matrix}\right],
$$
of $h_\cS$ for fixed $L$; see Proposition~\ref{prop:lowerbound2} below. 

The connection between the $m$-functions and the transfer matrix is provided 
by the following lemma (see~\cite[Lemma 3.3]{BJLP1}).
 
\begin{lemma}\label{lem:eigenfunction-mfunction} For any $E\in\sp(h_{\per,L})$,
the eigenvalues of $T(L,E)$ are of the form $\e^{\pm\i \theta(L,E)}$ and 
$$
\psi_+(L,E)=\left[\begin{array}{c}1\\m_r(L,E)^{-1}\end{array}\right]
\quad\mbox{and}\quad
\psi_-(L,E)=\left[\begin{array}{c}1\\m_l(L,E)\end{array}\right],
$$
are corresponding eigenvectors.
\end{lemma}

\noindent {\bf Remark.} The fact that the eigenvalues of $T(L,E)$ are complex 
conjugate further implies the following relation between the two 
$m$-functions: $m_r(L,E)^{-1}= \overline{m_l}(L,E)$, i.e. 
$m_r\overline{m_l}=1$.

The following proposition provides a lower bound of the CLB conductance in 
terms of the transfer matrix $T(L,E)$ and the imaginary part of its 
eigenvalues.
\bep\label{prop:lowerbound1} If there exists $\delta,M>0$ such that 
$\Im F_{l/r}(E+i0)>\delta$ and $|F_{l/r}(E+i0)|\leq M$ for a.e.\;$E\in I$, 
then there exists $C>0$ such that for any $L$
\beq\label{eq:lowerbound1}
G_{\rm CLB}(L,I)\geq\frac{1}{2\pi|I|}\int_{I\cap\sp(h_{\per,L})}
\left[1+C\frac{\|T(L,E)\|}{|\sin(\theta(L,E))|}\right]^{-1}\d E.
\eeq
\eep

\proof One easily gets from~\eqref{def:crys-tra-coef} that 
$$
G_{\rm CLB}(L,I)\geq\frac{1}{2\pi|I|}\int_{I\cap\sp(h_{\per,L})}
\left[1+\frac{A+B|m_r(L,E)|^2}{2\Im(m_r(L,E))}+ 
\frac{A+B|m_l(L,E)|^2}{2\Im(m_l(L,E))}\right]^{-1}\d E,
$$
with $A=\frac{\kappa^2M^2}{\delta}$ and $B=\frac1{\kappa^2\delta}$, where we 
used that $I\subset\Sigma_l\cap\Sigma_r$. Since $m_r\overline{m_l}=1$, 
$$
\frac{A+B|m_r(L,E)|^2}{\Im(m_r(L,E))}=\frac{A|m_l(L,E)|^2+ B}{\Im(m_l(L,E))}.
$$
Hence, with $C=\max(A,B)$, we have
\beq\label{eq:conductancelowerbound}
G_{\rm CLB}(L,I)\geq\frac{1}{2\pi|I|}\int_{I\cap \sp(h_{\per,L})}
\left[1+C\frac{1+|m_l(L,E)|^2}{\Im(m_l(L,E))}\right]^{-1}\d E.
\eeq

We now relate the integrand on the right-hand side of the last inequality to 
the transfer matrix $T(L,E)$. Using Lemma~\ref{lem:eigenfunction-mfunction} 
and $m_r\overline{m_l}=1$, an easy calculation gives
$$
T(L,E) = \frac1{\Im (m_l(L,E))}\left[\begin{array}{cc}
\Im (\e^{i\theta(L,E)}m_l(L,E))&-\sin(\theta(L,E))\\ 
|m_l(L,E)|^2\sin(\theta(L,E))&\Im (\e^{-i\theta(L,E)}m_l(L,E))  
\end{array}\right],
$$ 
from which we get the lower bound
$$
\|T(L,E)\| \geq C'|\sin(\theta(L,E))| \frac{1+|m_l(L,E)|^2}{\Im (m_l(L,E))},
$$
for some positive constant $C'$. Inserting this inequality 
into~\eqref{eq:conductancelowerbound} completes the  proof.
\hfill\qed

In view of~\eqref{eq:lowerbound1}, to get a lower bound of the CLB conductance 
in terms of the norm of the transfer matrix, the only issue is when 
$|\sin(\theta(L,E))|$ gets small. The following shows that this cannot happen 
too often. In the  sequel, $|A|$ denotes the Lebesgue measure of 
$A\subset \rr$. 

\bel\label{lem:theta-small} For any $\epsilon>0$ and all $L$, 
$$
|\left\{ E\in\sp(h_{\per,L})\,:\,|\sin(\theta(L,E))|\leq\epsilon\right\}|
\leq 2\pi\epsilon.
$$
\eel

The proof of this lemma relies on a general estimate on the so-called 
dispersion curves of the periodic operator $h_{\per,L}$ and requires 
additional notation and facts. We postpone it to 
Section~\ref{ssec:proof-lemma}.

Combining Proposition~\ref{prop:lowerbound1} with Lemma~\ref{lem:theta-small} 
and recalling the inequality $\|T(L,E)\|\geq 1$, we get
\bep\label{prop:lowerbound2} If there exists $\delta,M>0$ such that 
$\Im F_{l/r}(E+i0)>\delta$ and $|F_{l/r}(E+i0)|\leq M$ for a.e.\;$E\in I$, 
then there exists $C>0$ such that for any $\epsilon>0$ and  all $L$,
$$
G_{\rm CLB}(L,I)\geq\frac{1}{2\pi|I|} 
\left(1+C\epsilon^{-1}\right)^{-1}
\int_{I\cap(\sp(h_{\per,L})\setminus\Omega_{\epsilon,L})} 
\|T(L,E)\|^{-1}\d E,
$$
with $|\Omega_{\epsilon,L}|\leq 2\pi\epsilon$.
\eep

Our last ingredient is the following estimate on the norm of transfer matrices 
which was proven in~\cite[Section 5.3]{BJLP2}.
\bep\label{prop:transfermatrixbound} There exists a set 
$O_{\epsilon,L}\subset\sp(h_{\per,L})$ such that 
$|O_{\epsilon,L}|\leq(1+\pi)\epsilon$ and
$$
\|T(L,E)\|\leq\frac{8\pi}{\epsilon^2},\qquad 
\forall E \in\sp(h_{\per,L})\setminus O_{\epsilon,L}.
$$
\eep

We are now in position to finish the proof of Theorem~\ref{main-th-1}. 
For any $n>0$, let
$$
I_n:=\{E\in I\,:\,\Im F_{l/r}(E+i0)>1/n\mbox{ and }|F_{l/r}(E+i0)|\leq n\},
$$
and $I_n'=I\setminus I_n$. Obviously, since
$I\subset\Sigma_l\cap\Sigma_r$, $\ds\lim_{n\to\infty}|I_n'|=0.$

Using Proposition~\ref{prop:lowerbound2} together with 
Proposition~\ref{prop:transfermatrixbound} on $I_n$, one gets that for any 
$n$ and for any $\epsilon>0$, 
\begin{eqnarray*}
|I|G_{\rm CLB}(L,I)&\geq&|I_n|G_{\rm CLB}(L,I_n)\\
&\geq&|I_n|\left(1+C_n\epsilon^{-1}\right)^{-1}
\int_{I_n\cap(\sp(h_{\per,L})\setminus \Omega_{\epsilon,L})}
\|T(L,E)\|^{-1}\d E\\
& \geq&|I_n|\left(1+C_n\epsilon^{-1}\right)^{-1}\frac{\epsilon^2}{8\pi}
\left(|\sp(h_{\per,L})\cap I_n|-|\Omega_{\epsilon,L}|-|O_{\epsilon,L}|\right)\\
&\geq&|I_n|\left(1+C_n\epsilon^{-1}\right)^{-1} \frac{\epsilon^2}{8\pi}
\left(|\sp(h_{\per,L})\cap I|-|I_n'|-(1+3\pi)\epsilon\right),
\end{eqnarray*}
for some $C_n$ which does not depend on $\epsilon$ and $L$.

Suppose now that the sequence $(L_k)$ is such that 
$\ds\lim_{k\to\infty}G_{\rm CLB}(L_k,I)=0$. It follows that
$$
\limsup_{k\to\infty}|\sp(h_{\per,L_k})\cap I|\leq|I_n'|+(1+3\pi)\epsilon.
$$
Since this holds for any $\epsilon>0$ and $|I_n'|\to0$, this proves that 
$\ds\lim_{k\rightarrow \infty}|\sp(h_{\per,L_k})\cap I|=0$, and hence 
$\ds\lim_{k\to\infty}G_{\rm Th}(L_k,I)=0$.


\subsection{Proof of Lemma \ref{lem:theta-small}}\label{ssec:proof-lemma}

We first introduce some notation and recall a few basic facts about periodic 
operators, referring  the reader to~\cite[Chapter~5]{Si} for proofs and 
additional information.

For any $k\in \rr$ and $m\in\zz$ let
\[
H(k,m)=\left[
\begin{matrix}
v_{\rm per}(m+1) &-1&\cdots&0&-\e^{-\i kL}\\
-1&v_{\rm per}(m+2)&\cdots&0&0\\
\vdots&\vdots&\ddots&\vdots&\vdots\\
0&0&\cdots&v_{\rm per}(m+L-1)&-1\\
-\e^{\i kL}&0&\cdots&-1&v_{\rm per}(m+L)
\end{matrix}
\right],
\]
and denote by $E_1(k)\le\cdots\le E_L(k)$ the repeated eigenvalues of 
$H(k,0)$. The functions $\rr\ni k\mapsto E_\ell(k)$ are called the dispersion 
curves and will be the key object in the proof of Lemma~\ref{lem:theta-small}. 
They are $2\pi/L$-periodic and even. They are strictly monotone and real 
analytic on the interval $(0,\pi/L)$. Moreover, they satisfy
$$
E_L(0)>E_L\left(\frac{\pi}L\right)\ge E_{L-1}\left(\frac{\pi}{L}\right)
>E_{L-1}(0)\ge E_{L-2}(0)>\cdots
$$
This implies in particular that each $E_\ell(k)$ is a simple eigenvalue of 
$H(k,0)$ for $k\in(0,\pi/L)$. It follows that for each 
$\ell\in\{1,\ldots,L\}$ there is a unique real analytic function
$$
(0,\pi/L)\ni k\mapsto\vec u_\ell(k)
=(u_\ell(k,1),\ldots,u_\ell(k,L))^T\in\cc^L,
$$ 
such that $H(k,0)\vec u_\ell(k)=E_\ell(k)\vec u_\ell(k)$, $u_\ell(k,1)>0$ and
$\|\vec u_\ell(k)\|=1$. A bounded two-sided sequence 
$u_\ell(k)=(u_\ell(k,m))_{m\in\zz}$ is obtained by setting
\beq
u_\ell(k,j+nL)=\e^{\i knL}u_\ell(k,j),
\label{eq:eigenvectors}
\eeq
for any $j\in\{1,\ldots,L\}$ and $n\in\zz$. Then, for any $m\in\zz$,
$$
\vec u_\ell(k,m)=(u_\ell(k,m+1),\ldots,u_\ell(k,m+L))^T,
$$
is a normalized eigenvector of $H(k,m)$ for the eigenvalue $E_\ell(k)$.

It follows from Floquet theory that $E\in\sp(h_{\per,L})$ iff the eigenvalue 
equation
\beq\label{eq:stationaryschrodinger}
h_{\per,L} u = Eu
\eeq
has a non-trivial solution $u$ satisfying $u(n+L)=\e^{\i kL}u(n)$ for some 
$k\in\rr$ and all $n\in\zz$. This solution is called Bloch wave of energy
$E$ and $u$ is such a Bloch wave if and only if $E=E_\ell(k)$ for some $\ell$ 
and $(u(1),\ldots,u(L))^T$ is an eigenvector of $H(k,0)$ for $E_\ell(k)$.
In particular, for any $m$,
$$
\sp(h_{\per,L})=\bigcup_{k\in [0,\pi/L]}\sp(H(k))=\bigcup_{\ell=1}^L B_\ell,
$$
where $B_\ell$ is the closed interval with boundary points $E_\ell(0)$ and 
$E_\ell(\pi/L)$. The $B_\ell$ are called spectral bands of $h_{\per,L}$
and have pairwise disjoint interiors. $E$ is an interior point of $B_\ell$ 
iff $E=E_\ell(k)$ for some $k\in(0,\pi/L)$. Because $E_\ell$ is monotone such 
a $k$ is unique and we shall denote it $k(E)$. On each band $B_\ell$, the 
function $k(E)$ is thus a strictly monotone function whose image is 
$(0,\pi/L)$.

The characteristic polynomial of $H(k,m)$ satisfies
$$
\det(H(k,m)-z)= \tr(T(L,z))-2\cos(kL).
$$
As a consequence, $\sp(h_{\per,L})=\{E\,:\,|\tr(T(L,E))|\leq 2\}$ and, 
for any $E\in\sp(h_{\per,L})$, $k(E)$ is determined by the identity
\begin{equation*}\label{eq:dispersioncurve}
\tr(T(L,E))=2\cos(k(E)L).
\end{equation*}
Since ${\rm det}(T(L,E))=1$, one infers that for $E\in\sp(h_{\per,L})$ the 
eigenvalues of $T(L,E)$ actually are $\e^{\pm\i k(E)L}$, 
i.e., $\theta(L,E)=\pm k(E)L$. The proof of Lemma~\ref{lem:theta-small}
relies of the following general estimate.
 
\bep\label{prop:dispersionbound} For any $\ell=1,\ldots,L$ and 
$k\in\left(0,\frac{\pi}{L}\right)$ one has
$$
|E'_\ell(k)|\leq 2.
$$
\eep

Although not explicitly stated, this result already appears in~\cite{BJLP2} 
(in the proof of Proposition~5.2). For the convenience of the reader we give 
its proof. 

\proof For any $k$ and $m$ the vector 
$\vec u_\ell(k,m)=(u_\ell(k,m+1),\ldots,u_\ell(k,m+L))^T$ is a normalized 
eigenvector of $H(k,m)$ for $E_\ell(k)$. The Feynman-Hellmann formula gives
\begin{align*}
E_\ell'(k)&=
\left\langle\vec u_\ell(k,m),
\frac{\d H(k,m)}{\d k}\vec u_\ell(k,m)\right\rangle\\
&=\i L\left(\overline{u_\ell(k,m+1)}\e^{-\i kL} u_\ell(k,m+L)
-\overline{u_\ell(k,m+L)}\e^{\i kL}u_\ell(k,m+1)\right),
\end{align*}
and the relation~\eqref{eq:eigenvectors} yields
\begin{equation*}\label{eq:derivativeEk}
E_\ell'(k)=2L\,\Im\left(\overline{u_\ell(k,m)}u_\ell(k,m+1)\right),
\end{equation*}
for all $m$. Summing over $m=1,\ldots,L$, we can write
$$
E_\ell'(k)=\sum_{m=1}^L 2\,\Im\left(\overline{u_\ell(k,m)}u_\ell(k,m+1)\right).
$$
The normalization of $u_\ell$ yields
$$
|E_\ell'(k)|
\leq\sum_{m=1}^L\left(|u_\ell(k,m)|^2 + |u_\ell(k,m+1)|^2\right)
=\|\vec u_\ell(k,0)\|^2+\|\vec u_\ell(k,1)\|^2=2.
$$
\hfill\qed

\noindent {\bf Proof of Lemma \ref{lem:theta-small}.} Fix $\epsilon>0$. Since 
$\theta(L,E)=\pm k(E)L$ is a monotone function in each spectral band $B_\ell$ 
and varies between $0$ and $\pi$, $|\sin(\theta(L,E))|\leq \epsilon$ can only 
hold near the band edges ($\sin(\theta(L,E))$ vanishes precisely at these edges). 
From 
Proposition~\ref{prop:dispersionbound} we get that for any 
$E\in \sp(h_{\per,L})$
\beq\label{eq:theta-estimate}
|\theta'(L,E)| \geq \frac{L}{2}.
\eeq
Using $\sin(\alpha)\geq\frac{2\alpha}{\pi}$ for 
$0\leq \alpha\leq \frac{\pi}{2}$ or $\sin(\alpha)\geq2-\frac{2\alpha}{\pi}$ 
for $\frac{\pi}{2}\leq \alpha\leq\pi$ one has, for any band $B_\ell$, 
$$
\left|\left\{E\in B_\ell\,:\,|\sin(\theta(L,E))|\leq \epsilon\right\}\right| 
\leq\left|\left\{E\in B_\ell\,:\,
|\theta(L,E)|\leq\frac{\epsilon\pi}{2}\right\}\right|
+\left|\left\{E\in B_\ell\,:\,|\pi-\theta(L,E)|\leq 
\frac{\epsilon\pi}{2}\right\}\right|,
$$
which together with~\eqref{eq:theta-estimate} gives
$$
\left|\left\{ E \in B_\ell\,:\,|\sin(\theta(L,E))|\leq \epsilon\right\}\right| 
\leq \frac{2\epsilon\pi}{L}.
$$
Summing these inequalities over the $L$ bands  proves the lemma.
\hfill\qed

\end{document}